\documentstyle[aps,eqsecnum,epsfig]{revtex} 
\begin{document} 
\tighten 
\draft  
\preprint{hep-ph/9810393,PITT-98-07, LPTHE-98/46, NDHU-TH-98-01} 
%\twocolumn[\hsize\textwidth\columnwidth\hsize\csname@twocolumnfalse\endcsname
\title{\bf Fermion Damping in a Fermion-Scalar Plasma}
\author{\bf D. Boyanovsky$^{(a)}$, H.J. de Vega$^{(b)}$, 
D.-S. Lee$^{(c)}$, Y.J. Ng$^{(d)}$, S.-Y. Wang$^{(a)}$}
\address
{(a) Department of Physics and Astronomy, University of Pittsburgh, 
Pittsburgh  PA. 15260, U.S.A\\ 
(b) LPTHE, Universit\'e Pierre et Marie Curie (Paris VI) et Denis
Diderot (Paris VII), Tour 16, 1er. \'etage, 4, Place Jussieu, 
75252 Paris, Cedex 05, France\\
(c) Department of Physics, National Dong Hwa University, Shoufeng, 
Hualien 974, Taiwan, R.O.C.\\
(d)Department of Physics and Astronomy, University of North Carolina, 
Chapel Hill, N.C. 27599, U.S.A.}
\date{October 16, 1998}
\maketitle 

\begin{abstract} 
In this article we study the dynamics of fermions
 in a fermion-scalar plasma. We begin by obtaining the effective
in-medium Dirac equation in real time which is fully renormalized and
 causal and  leads to the initial value problem. For a heavy scalar we
 find the novel result that the {\em decay} of the scalar into fermion
 pairs in the medium leads to damping of the fermionic excitations and
 their in-medium propagation as quasiparticles. That is, the fermions
 acquire a width due to the decay of the heavier scalar in the medium.
We find the damping rate to lowest order in the Yukawa coupling for
 arbitrary values of scalar and fermion masses, temperature and
 fermion momentum. An all-order expression for the damping rate in
 terms of the exact quasiparticle  wave functions is established. 
A kinetic Boltzmann approach to the relaxation of the fermionic distribution 
function confirms the damping of fermionic excitations as a
 consequence of the induced decay of heavy scalars in the medium.  
A linearization of the Boltzmann equation near equilibrium clearly 
displays the relationship between the damping rate of fermionic mean
 fields and the fermion interaction rate to lowest order in the Yukawa coupling 
directly in real time.  
\end{abstract} 
%\vspace{2pc}]

\section{Introduction} 
The propagation of quarks and leptons in a medium of high temperature and/or 
density is of fundamental importance in a wide variety of physically relevant 
situations. In stellar astrophysics, electrons and neutrinos
play a major role in the evolution of dense stars such as white dwarfs, 
neutron stars and supernovae~\cite{raffelt}. 
In ultrarelativistic heavy ion collisions and the possibility of formation of a 
quark-gluon plasma, electrons (and muons) play a very important role as 
clean probes of the early, hot stage of the plasma~\cite{rhic}. 
Furthermore, medium effects can enhance neutrino oscillations as envisaged
in the Mikheyev-Smirnov-Wolfenstein (MSW) effect~\cite{wolf} and 
dramatically modify the neutrino electromagnetic couplings~\cite{woo}. 
The propagation of quarks during the non-equilibrium stages of the 
electroweak phase transition is conjectured to be an essential ingredient for 
baryogenesis at the electroweak scale both in 
non-supersymmetric and supersymmetric extensions of the standard model~\cite{baryo}. 

In-medium propagation is dramatically different from that in vacuum. 
The medium modifies the dispersion relation of the excitations and introduces 
a width to the propagating excitation~\cite{weldon1,weldon2,kapusta,lebellac} 
that results in damping of the amplitude of the propagating mode.
In this article we focus on several aspects of propagation of fermionic
excitations in a fermion-scalar plasma: 

\begin{itemize}
\item We begin by deriving the effective and fully renormalized 
Dirac equation {\em in real time}. 
This is achieved by relating the expectation
value of a fermionic field induced by an external fermionic source via
linear response  
to an initial value problem for the expectation value. 
This initial value problem is in terms of the effective real time
Dirac equation in the medium that is i) renormalized, ii) retarded and causal. 

The necessity for a consistent Dirac equation in a medium has been recognized 
in the literature within the context of neutrino oscillations in the 
medium~\cite{dolivo1,weiss,kim,widom}. 
In particular, in references~\cite{dolivo1,weiss} a proposal for 
a field-theoretical treatment of neutrino oscillations in the medium 
starting from the Dirac equation was presented.

In this article we introduce the fully renormalized, in-medium
effective Dirac equation in {\em real time} that allows a more transparent
study of damping and oscillations in time. There are definite advantages 
in such formulation since the real time evolution is obtained at once from
an initial value problem and allows a straightforward identification of
the damping rate.   

\item We apply the effective Dirac equation in a medium to study the 
real-time evolution of fermionic excitations in a fermion-scalar plasma. 
Whereas the propagation of quarks and leptons in a QED or QCD plasma has been 
studied thoroughly (see ref.~\cite{kapusta,lebellac} for details)
a similar study for a scalar plasma has not been carried out to the
same level of detail.  
Recently some attention has been given to understanding the
thermalization time scales  
of bosonic and fermionic
excitations in a plasma of gauge~\cite{vilja1} 
and scalar bosons~\cite{vilja2}, 
furthermore fermion thermalization is an important ingredient 
in models of baryogenesis mediated by scalars~\cite{krishna}. 
Most of the studies of fermion thermalization focus on
the mechanism of fermion scattering off the gauge quanta 
in the heat bath and/or Landau damping in the hard thermal
loop (HTL) resummation program~\cite{htl}. 
Although the scalar contribution to the fermionic self-energy to one
loop has been obtained a long
time ago~\cite{weldon2}, scant attention has been paid to a more detailed 
understanding of the contribution from the scalar degrees of freedom to the
fermion relaxation and thermalization. 
As mentioned above this issue becomes of pressing importance in models of 
baryogenesis and more so in models in which the scalars carry baryon
number~\cite{krishna}.  

Whereas the contribution from scalars to the fermionic thermalization
time scale (damping rate) has been studied  for massless chiral
fermions~\cite{dolivo1,dolivodamp1},  
in this article we offer a detailed and general study of fermion relaxation 
and thermalization through the interactions with the scalars in the
plasma in real time and for arbitrary values of the scalar and fermion masses, 
temperature and fermion momentum. 
More importantly, we focus on a novel mechanism of damping of fermionic 
excitations that occurs whenever the effective mass of the scalar 
particle allows its kinematic {\em decay} into fermion pairs. 
This phenomenon only occurs in a medium and is
interpreted as an induced decay of the scalar in the medium. 
It is a process different from collisions with particles in the bath 
and Landau damping which are the most common processes that lead to 
relaxation and thermalization.

This process results in new thermal cuts in the
fermionic self-energy and for heavy scalars, this cut results in a
quasiparticle pole  structure for the fermion and provides a width for
the fermionic quasiparticle.  
The remarkable and perhaps non-intuitive aspect of this 
process is that the decay of the scalar results in {\em damping} of the
fermionic excitations and their propagation as  quasiparticle
resonances. Our real time analysis reveals that amplitude of
 the $ {\vec k} $-mode of the expectation value of
the fermion field decreases as $ e^{- \Gamma_k\, t} $ while it oscillates with 
frequency $\omega_{\text{p}}(k)$ which is determined by the position of the resonance.

The effective real-time Dirac equation in the medium allows a direct 
interpretation of the damping of the quasiparticle fermionic excitation and 
leads to a clear definition of
the damping rate. By analyzing the quasiparticle wave functions we obtain an 
{\em all-order} expression for the damping 
rate $ \Gamma_k $ and confirm and generalize recent results for the
massless chiral case~\cite{dolivo1,dolivodamp1}.  

\item In order to provide a complementary understanding of the process
of induced decay of the heavy scalars in the medium and the resulting
fermion damping,  we study the kinetics of relaxation of the fermionic
distribution function via a Boltzmann equation to lowest order.  
Linearizing the Boltzmann equation near the equilibrium  distribution, 
we obtain the relation between the thermalization rate for the
distribution function in the relaxation time approximation (linearized
near equilibrium) and the damping rate for the amplitude of the
fermionic mean fields to lowest order in the Yukawa coupling. This
analysis provides a real-time confirmation of the oft quoted relation
between the interaction rate (obtained from the Boltzmann kinetic equation in
the relaxation time approximation) and the damping rate for the mean 
field~\cite{weldon2,lebellac}.
More importantly, this analysis reveals directly, via a kinetic approach 
in real time how the process of induced decay of a heavy scalar in the medium 
results in damping and thermalization of the fermionic excitations. 
A study of the relation between the interaction rate and the damping rate has
been presented recently for gauge theories within the context of the
imaginary time formulation~\cite{dolivodamp2}. Our results provide a real-time 
confirmation of those of reference~\cite{dolivodamp2} for the scalar
case.  
\end{itemize}

The article is organized as follows: in section II we obtain the effective 
in-medium Dirac equation in real time starting from the linear response to 
an external Grassmann-valued source that induces a mean field. 
We obtain the {\em fully renormalized}  Dirac equation with the real
time self-energy to one loop order by turning the linear response
problem into an initial value problem for the mean field. The
renormalization aspects are addressed in detail in this section. In
section III we study in detail the structure of the renormalized
self-energy and  establish the presence of new cuts of thermal origin. 
We then note that for heavy scalars such that their decay into fermion
pairs is kinematically allowed, the fermionic pole becomes embedded in
this thermal cut resulting in a quasiparticle (resonance) structure,
which is analyzed in detail. The decay rate is evaluated in the narrow
width approximation (justified for small Yukawa couplings)
 for {\em arbitrary values}  
of the scalar and fermion masses, temperature and fermion momentum. 

In section IV we present a real time analysis of the evolution of the
mean-fields. In this section we clarify the difference between complex
poles and resonances (often misunderstood). This analysis reveals
clearly that the induced decay of the scalar results in an exponential
damping of the amplitude of the mean field and yields to a clear
identification of the damping rate bypassing the conflicting definitions  
of the damping rate offered in the literature. An analysis of the
structure of the self-energy and an interpretation of the {\em exact}
quasiparticle spinor wave functions   
allows us to provide an {\em all-order} expression for the damping rate of
the fermionic mean fields. In section V we present an analysis of the
evolution of the distribution functions in real time by obtaining a Boltzmann
kinetic equation for the spin-averaged fermionic distribution function. 
In the linearized approximation near equilibrium (relaxation time
approximation), we clarify to lowest order the relation between the damping
rate of the quasiparticle fermionic excitations and the thermalization rate 
(linearized) of the distribution function, directly in real time. 
In the conclusions we summarize our results and suggest new avenues including 
the contribution from gauge fields. In this section we also
assess the importance of the scalar contribution in theories with gauge
and scalar fields. 

\section{Effective Dirac equation in the medium}

As mentioned in the introduction, whereas the damping of collective and 
quasiparticle excitations via the interactions with gauge bosons in
the medium has been the focus of most attention, understanding of the
influence of scalars has not been pursued  so vigorously. 

Although we are ultimately interested in studying the damping of 
fermionic excitations in a plasma with scalars  and gauge fields within
the realm of electroweak baryogenesis in either the Standard Model or
generalizations
thereof, we will begin by considering only the coupling of a massive
Dirac fermion  to a scalar via a simple Yukawa interaction. 

The model dependent generalizations of the Yukawa couplings to particular 
cases will differ quantitatively
in the details of the group structure but the qualitative features of
the effective Dirac equation in the medium as well as the kinematics
of the thermal cuts that lead to damping of the fermionic excitations
will be rather general.  

We consider a Dirac fermion with the bare mass $M_0$ coupled to a
scalar with the bare mass $m_0$ via a Yukawa coupling. The bare fermion
mass could be the result of spontaneous symmetry breaking in the
scalar sector, but for the purposes of our studies we need not specify
its origin.  

The Lagrangian density is given by
\begin{eqnarray}
&&{\cal{L}}= \bar{\Psi}(i{\not\!{\partial}}-M_0)\Psi + 
\frac{1}{2} \partial_\mu\phi\,\partial^\mu\phi-
\frac{1}{2} m_0^2 \phi^2- {\cal L}_I[\phi] -
y_0\bar{\Psi}\phi\Psi+\bar{\eta}\Psi+\bar{\Psi}\eta+j\phi~,
\label{yukalan}
\end{eqnarray}
where  $y_0$ is the bare Yukawa coupling. The self-interaction of the scalar
field accounted for by the term ${\cal L}_I[\phi]$ need not be specified to lowest order. 
The $\eta$ and $j$ are the respective external fermionic and 
scalar sources that are introduced in order to provide an initial value 
problem for the effective Dirac equation.
We now write the bare fields and sources 
$\Psi$, $\phi$, $\eta$ and $j$ in terms of the renormalized quantities
(hereafter referred to with a subscript $r$) by introducing the renormalization 
constants and counterterms:
\begin{eqnarray}
&&\Psi=Z_\psi^{1/2}~\Psi_r~,\;\phi=Z_\phi^{1/2}~\phi_r ~,\;
\eta=Z_{\psi}^{-1/2}\eta_r ~,\;j=Z_\phi^{-1/2}j_r~, \nonumber \\
&& y=y_0 Z_\phi^{1/2} 
Z_\psi / Z_y ~,\; m^2_0  = \left(\delta_m +m^2\right)/Z_{\phi}~,\; 
M_0 = \left(\delta_M +M\right)/Z_{\psi}~.  
\label{wfren}
\end{eqnarray}
With the above definitions, the ${\cal{L}}$ can be expressed as:
\begin{eqnarray}
{\cal{L}}&=&\bar{\Psi}_r (i{\not\!{\partial}}-M)\Psi_r 
+ \frac{1}{2} \partial_\mu\phi_r\,\partial^\mu\phi_r-
\frac{1}{2} m^2 \phi_r^2- {\cal L}_I^r[\phi_r]- y \bar{\Psi}_r\phi_r\Psi_r 
 + \bar{\eta}_r\Psi_r 
+\bar{\Psi}_r\eta_r+j_r \phi_r\nonumber\\
&& + \frac{1}{2} \delta_\phi \partial_\mu\phi_r\,\partial^\mu\phi_r -
\frac{1}{2}
\delta_m \phi_r^2 
+\bar{\Psi}_r(i\delta_\psi{\not\!{\partial}}-\delta_M)\Psi_r
 -y \delta_y \bar{\Psi}_r\phi_r\Psi_r+ \delta 
{\cal L}^r_{I}[\phi_r]~,\label{renyukalan}
\end{eqnarray}
where $m$ and $M$ are the renormalized masses, and 
$y$ is the renormalized Yukawa coupling. The terms with the coefficients
\begin{eqnarray}
&&\delta_\psi=Z_\psi-1~,\;\delta_\phi=Z_\phi-1~,\nonumber\\
&&\delta_M= M_0 Z_\psi-M~,\;\delta_m= m^2_0 Z_\phi-m^2~,\\
&&\delta_y=Z_y-1 \nonumber
\end{eqnarray}
and $\delta {\cal L}^r_I$ are the counterterms to be 
determined consistently in the perturbative
expansion by choosing a renormalization prescription.  As it will become
clear below this is the most natural manner for obtaining a fully
renormalized Dirac equation in a perturbative expansion.  

The dynamics of expectation values and correlation functions of the
 quantum field is obtained by
implementing the Schwinger-Keldysh closed-time-path formulation of non-equilibrium 
quantum field theory~\cite{ctp,disip,tadpole,linon}.  
The main ingredient in this formulation is the {\em real} time evolution of an
initially prepared density matrix and its path integral
representation. It requires a path integral defined along a closed
time path contour.  This formulation has been
described elsewhere within many different contexts and we refer the 
reader to the literature for details~\cite{ctp,disip,tadpole,linon}. 

Our goal is to understand the non-equilibrium relaxational dynamics of 
the inhomogeneous fermionic mean fields 
$$ 
\psi(\vec{x},t) \equiv \langle \Psi_r(\vec{x},t)\rangle 
$$
from their initial states in the presence of the fermion-scalar medium. 

This statement requires clarification. In states  of definite fermion number (either zero or
finite temperature) the expectation value of the fermion field must necessarily vanish. However, 
in order to understand the non-equilibrium dynamics, we prepare the system by coupling an 
external Grassman source to the fermionic field in the Hamiltonian. This source creates a
coherent state of fermions which is a superposition of states with different fermion number. 
As will be discussed in detail below, this source term is switched-on adiabatically from $t=-\infty$ 
in the {\em full interacting theory} up to
a time $t_0$ and switched-off at $t=t_0$. 
The resulting fermionic coherent state at $t=t_0$ does not have a definite fermion number and the expectation
value of the field operator in this state is non-vanishing. After the source is switched-off at 
time $t_0$  the expectation value evolves in time with the full interacting Hamiltonian in
absence of sources.  This is the standard approach to linear response, in this approach the source term
couples to a given mode of wavevector $k$ of the fermionic fields, thus displacing this degree of freedom
off-equilibrium. The other modes are assumed to be remain in thermal equilibrium.

Since we are interested in real time correlation functions and the
initial density matrix is assumed to be that of thermal equilibrium 
at initial temperature $ T=1/\beta $ with respect to the free
(quadratic) Lagrangian,   
only the real time branches, forward and backwards are required. The
contribution from the imaginary time branch corresponding to the thermal
component of the density matrix cancels in the connected non-equilibrium
expectation values~\cite{disip,tadpole,linon}. The effective non-equilibrium 
Lagrangian density that enters in the contour path integral is
therefore given by  
\begin{equation}
{\cal L}_{\text{non-eq}}= {\cal
L}\left[\Psi_r^+,\bar{\Psi}_r^+,\phi_r^+\right]
- {\cal
L}\left[\Psi_r^-,\bar{\Psi}_r^-,\phi_r^-\right]\;.\label{noneqlag}
\end{equation}
Fields  with $(+)$ and $(-)$ superscripts are defined respectively 
on the forward ($+$) and backwards ($-$) time contours 
and are to be treated independently. The external sources are the same
for both branches. 

The essential ingredients for perturbative calculations are the
following
real time Green's functions~\cite{linon}:

\begin{itemize}
\item{Scalar Propagators
\begin{eqnarray}
&& G_{\vec{k}}^{++}(t,t') = G_{\vec{k}}^{>}(t,t')\Theta(t-
t')+G_{\vec{k}}^{<}(t,t')\Theta(t'-t)\;,
\label{timeord}  \nonumber \\
&& G_{\vec{k}}^{--}(t,t') = G_{\vec{k}}^{>}(t,t')\Theta(t'-
t)+G_{\vec{k}}^{<}(t,t')\Theta(t-t')\;,
\label{antitimeord} \nonumber \\
&& G_{\vec{k}}^{+-}(t,t')= - G_{\vec{k}}^{<}(t,t') \label{plusmin} \;, \nonumber \\
&& G_{\vec{k}}^{-+}(t,t')= - G_{\vec{k}}^{>}(t,t') \label{minplus} \;, \nonumber \\
&&G_{\vec{k}}^{>}(t,t')= i \int d^3x  \; e^{-i\vec{k}\cdot\vec{x}} \;
\langle \phi_r(\vec{x},t) \phi_r(\vec{0},t') \rangle \nonumber \\
&&\quad\quad\quad~~=\frac{i}{2\omega_k}\left[(1+n_k)
e^{-i\omega_k(t-t')}+n_k e^{i\omega_k(t-
t')}\right]\;,
\label{ggreat} \nonumber  \\
&&G_{\vec{k}}^{<}(t,t')= i \int d^3x \; e^{-i\vec{k}\cdot\vec{x}} \;
\langle \phi_r(\vec{0},t') \phi_r(\vec{x},t) \rangle \nonumber \\
&&\quad\quad\quad~~=\frac{i}{2\omega_k}\left[n_k
e^{-i\omega_k(t-t')}+(1+n_k)e^{i\omega_k(t-t')}\right]\;,
\label{gsmall} \nonumber  \\
&&\quad \omega_k=\sqrt{\vec{k}^2+m^2}\;,\quad\quad\quad  n_k=
\frac{1}{e^{\beta \omega_k}-1} \;.\label{bosonprops} 
\end{eqnarray}}

\item{Fermionic Propagators (Zero chemical potential)
\begin{eqnarray}
&& S_{\vec{k}}^{++}(t,t')=S_{\vec{k}}^{>}(t,t')\Theta(t-t')
+S_{\vec{k}}^{<}(t,t')\Theta(t'-t) \;,\nonumber\\
&& S_{\vec{k}}^{--}(t,t')=S_{\vec{k}}^{>}(t,t')\Theta(t'-t)
+S_{\vec{k}}^{<}(t,t')\Theta(t-t') \;,\nonumber\\
&& S_{\vec{k}}^{+ -}(t,t')=-
S_{\vec{k}}^{<}(t,t')\;,\nonumber\\
&& S_{\vec{k}}^{- +}(t,t')=-
S_{\vec{k}}^{>}(t,t')\;,\nonumber\\
&& S_{\vec{k}}^{>}(t,t')=-i\int d^3x \; e^{-i\vec{k}\cdot
\vec{x}} \; \langle \Psi_r(\vec{x},t) \bar{\Psi}_r(\vec{0},t')
\rangle  \nonumber\\
&&\quad\quad\quad~~=-\frac{i}{2\bar{\omega}_k}\left[
(\not\!{k}+M) (1-\bar{n}_k)e^{-i\bar{\omega}_k(t-t')}+
\gamma_0 (\not\!{k}-M)\gamma_0 \bar{n}_k
e^{i\bar{\omega}_k(t-t')}\right]~, \nonumber \\
&& S_{\vec{k}}^{<}(t,t')=i\int d^3x \; e^{-i\vec{k}\cdot
\vec{x}} \;  \langle \bar{\Psi}_r(\vec{0},t')  \Psi_r(\vec{x},t)
\rangle \nonumber \\
&&\quad\quad\quad~~=\frac{i}{2\bar{\omega}_k}\left[(\not\!{k}+M)
\bar{n}_k e^{-i\bar{\omega}_k(t-t')}
+\gamma_0
(\not\!{k}-M)\gamma_0 (1-\bar{n}_k) e^{i\bar{\omega}_k(t-t')}\right]\;,
\label{slessdef}\nonumber\\
&&\quad \bar{\omega}_k=\sqrt{\vec{k}^2+M^2}\;,\quad\quad\quad
\bar{n}_k=\frac{1}{e^{\beta \bar{\omega}_k}+1}\;. \label{fermionprops}
\end{eqnarray}}
\end{itemize}

The perturbative evaluation of correlation functions proceeds as usual,
but now the Feynman rules involves two types of vertices with opposite
signs and the four different non-equilibrium propagators. The symmetry factors
are the usual ones.  

These free propagators (\ref{gsmall}) and
(\ref{fermionprops}) are thermal since the initial state is assumed to be in thermal equilibrium.

\subsection{In-medium Dirac equation from linear response}

Consider the fermionic mean field obtained as the linear response to 
the externally applied (Grassmann-valued) source $\eta_r$: 
\begin{eqnarray}
&&\langle \Psi_r^+(\vec{x},t)\rangle = 
\langle \Psi_r^-(\vec{x},t)\rangle = \psi(\vec{x},t) =
-\int_{-\infty}^{\infty} dt' d^3x'
S_{\text{ret}}(\vec{x}-\vec{x}',t-t')\;  \eta_r(\vec{x}',t')~, \label{expecval}
\end{eqnarray}
with the {\em exact} retarded Green's function
\begin{eqnarray}
S_{\text{ret}}(\vec{x}-\vec{x}',t-t')&=&
\left[S^>(\vec{x}-\vec{x}',t-t')-S^<(\vec{x}-\vec{x}',t-t')\right]
\Theta(t-t')\nonumber\\
&=&-i\langle\{\Psi_r(\vec{x},t),\bar{\Psi}_r(\vec{x}',t')\}\rangle
\Theta(t-t')~,
\end{eqnarray}
where the expectation values are in the full interacting theory but
with vanishing sources. 
An initial value problem is obtained by considering that 
the external fermionic sources  are adiabatically 
switched on in time from $t\rightarrow -\infty$ thereby inducing an
expectation value of the fermionic fields, and switching-off the
source term at some time $t_0$. Then for $t>t_0$ this expectation value
or
mean field will evolve in the absence of a source and will relax because
of the interactions. The evolution for $t>t_0$ is an initial value
problem,
since the source term was used to prepare an initial state and switched
off
to let this state evolve in time. This initial value problem can
therefore
be formulated by choosing the source term to be of the form
\begin{eqnarray}
\eta_r(\vec{x},t)&=&\eta_r(\vec{x})\; e^{\epsilon t }\; \Theta(t_0-t)~r,
\quad \epsilon\rightarrow 0^{+}~.
\label{source}
\end{eqnarray}

In what follows we choose $t_0=0$ for convenience. The adiabatic
switching-on of the source induces an expectation
value that is dressed adiabatically by the interaction. 
The retarded and the equilibrium nature of
$S_{\text{ret}}(\vec{x}-\vec{x}',t-t')$  
(which depends on the time difference) and the
form of the source (\ref{source}) guarantee that 
\begin{eqnarray}
&&\psi(\vec{x},t=0)=\psi_{0}(\vec{x})~,\nonumber \\
&&\dot{\psi}(\vec{x},t<0)=0~\; ,\nonumber
\end{eqnarray}
where $\psi_{0}(\vec{x})$ is determined by 
$\eta_r(\vec{x})$ (or vice versa, the initial conditions for the
condensates $\psi_{0}(\vec{x})$ can be used to find $\eta_r(\vec{x})$). This
can be seen by taking the time derivative of $\psi(\vec x,t)$ in
eq.~(\ref{expecval}),  
using the fact that the retarded propagator depends on
the time difference, integrating by parts, and using the form of the 
external source and the retarded nature of the propagator. 

In order to relate this linear 
response problem to the initial value problem for the dynamical 
equation of the mean field, let us consider the (integro-) differential
operator ${\cal{O}}_{( {\vec{x}},t)}$ which is the  inverse of
$S_{\text{ret}}(\vec{x}- \vec{x}',t-t')$
so that 
\begin{equation}
{\cal{O}}_{( {\vec{x}},t)}\psi(\vec{x},t)=-\eta_{r}(\vec{x},t)~,\quad
\psi(\vec{x},t=0)=\psi_{0}(\vec{x})~,\quad \dot{\psi}(\vec{x},t<0)=0~,
\end{equation}
where the source is given by eq.~(\ref{source}).

It is at this stage where the non-equilibrium formulation provides the
most powerful framework. The real-time equations of motion for the mean
fields can be obtained via the tadpole method~\cite{disip,tadpole,linon}, which
automatically leads
to a retarded and causal initial value problem for the expectation value of the field. 
The implementation of this method
is as follows. Let us  introduce the inhomogeneous mean fields 
$\psi(\vec{x},t) = \langle \Psi(\vec x,t)\rangle $ and $
\bar{\psi}(\vec{x},t) =\langle \bar{\Psi}(\vec x,t)\rangle $. 
The dynamics
of these fermionic mean fields  in the plasma can  
be analyzed by treating 
$\psi(\vec{x},t)$ and $\bar{\psi}(\vec{x},t)$ as
background fields, i.e., the expectation values of the 
corresponding fields in the non-equilibrium density matrix, by expanding
the non-equilibrium Lagrangian density  about these mean fields. 
Therefore, we write the 
full quantum fields as the $c$-number expectation values
(mean fields) and quantum fluctuations about
them:
\begin{eqnarray}
&\Psi_r^{\pm}(\vec{x},t)=\psi(\vec{x},t)
+\chi^{\pm}(\vec{x},t)~,\quad
\bar{\Psi}_r^{\pm}(\vec{x},t)=\bar{\psi}(\vec{x},t)
+\bar{\chi}^{\pm}(\vec{x},t)~,&
%\label{condensate1}\\
\nonumber \\
&\psi(\vec{x},t)=\langle\Psi_r^{\pm}(\vec{x},t)\rangle~, \quad
\bar{\psi}(\vec{x},t)=\langle\bar{\Psi}_r^{\pm}(\vec{x},t)\rangle~,&
\label{condensate2}
\end{eqnarray}
where the expectation values of the field operators are taken in the
time-evolved density matrix.

The equations of motion for the mean fields can be 
obtained to any order
in the perturbative expansion by imposing the requirement that
the expectation value of the quantum fluctuations in the time evolved
density matrix vanishes identically.  This is referred to
as the tadpole equation~\cite{disip,tadpole,linon} which  follows  from
 eqs.~(\ref{condensate2}):
\begin{equation}
\langle \chi^{\pm}\rangle=0~, \quad 
\langle \bar{\chi}^{\pm}\rangle=  0~. \label{tadpoleq}
\end{equation}
The procedure consists in treating the {\em linear} terms in
$\chi^{\pm}$ as well as the non-linearities in perturbation theory and
keeping only the
connected irreducible diagrams, as in usual perturbation theory.
The equations obtained via this procedure are the equations of motion
obtained by variations of the non-equilibrium effective 
action~\cite{ctp,disip,tadpole,linon}.

By applying the tadpole method and taking spatial Fourier transform,
we find the effective {\em real time} Dirac equation  for the $ {\vec
k}$-mode of the expectation value  of the fermion: 
\begin{eqnarray}
&&\left[\left(i\gamma_0\frac{\partial}{\partial t} 
- \vec{\gamma}\cdot\vec{k}-M\right)+
\delta_\psi\left(i\gamma_0\frac{\partial}{\partial t}
- \vec{\gamma}\cdot\vec{k}\right)
-\delta_M\right] \psi_{\vec{k}}(t)+
\int_{-\infty}^t dt' \; \Sigma_{\vec{k}}(t-t')\; \psi_{\vec{k}}(t')= -
\eta_{r,\vec{k}}(t)~, \nonumber
\end{eqnarray}
where $\Sigma_{\vec{k}}(t-t')$ is the fermion self-energy and
$$
\psi_{\vec{k}}(t) \equiv \int d^3x\;  e^{-i \vec{k} \cdot \vec{x} } \;
\psi(\vec{x},t) \;\;.
$$
Using the non-equilibrium Green's functions
(\ref{bosonprops}) and (\ref{fermionprops}), we find to one loop
order, that $\Sigma_{\vec{k}}(t-t')$ is given by
\begin{eqnarray}
&&\Sigma_{\vec{k}}(t-t')=i\gamma_0\;\Sigma^{(0)}_{\vec{k}}(t-t')+
\vec{\gamma}\cdot\vec{k}~~\Sigma^{(1)}_{\vec{k}}(t-t')+
\Sigma^{(2)}_{\vec{k}}(t-t')~\;\;,
\end{eqnarray}
where
\begin{eqnarray}
\Sigma^{(0)}_{\vec{k}}(t-t')&=& y^2 \int \frac{d^3 q}{(2\pi)^3
}\frac{\bar{\omega}_{q}}{
2 \omega_{k+q} \bar{\omega}_{q}} \times\nonumber\\
&&\Bigl[\cos[(\omega_{k+q}+\bar{\omega}_q)(t-t')](1+n_{k+q}-\bar{n}_q)+
\cos[(\omega_{k+q}-\bar{\omega}_q)(t-t')](n_{k+q}+\bar{n}_q)\Bigr]~,\nonumber\\
\Sigma^{(1)}_{\vec{k}}(t-t')&=&  y^2 \int \frac{d^3 q}{(2\pi)^3
} \frac{1}{ 
2 \omega_{k+q} \bar{\omega}_{q}}\frac{\vec k \cdot \vec q}{k^2}
\times\nonumber\\ 
&&\Bigl[\sin[(\omega_{k+q}+\bar{\omega}_q)(t-t')](1+n_{k+q}-\bar{n}_q)-
\sin[(\omega_{k+q}-\bar{\omega}_q)(t-t')](n_{k+q}+\bar{n}_q)\Bigr]~,\nonumber\\
\Sigma^{(2)}_{\vec{k}}(t-t')&=& y^2 \int \frac{d^3 q}{(2\pi)^3
} \frac{M}{
2 \omega_{k+q} \bar{\omega}_{q}} \times\nonumber\\
&&\Bigl[\sin[(\omega_{k+q}+\bar{\omega}_q)(t-t')](1+n_{k+q}-\bar{n}_q)-
\sin[(\omega_{k+q}-\bar{\omega}_q)(t-t')](n_{k+q}+\bar{n}_q)\Bigr]~,\nonumber
\end{eqnarray}
with $\omega_{k+q}=\sqrt{(\vec{k}+\vec{q})^2+m^2}$ and
$n_{k+q}=(e^{\beta\omega_{k+q}}-1)^{-1}$ being, respectively,
the energy and the distribution function for scalars of momentum
$\vec{k}+\vec{q}$.

As mentioned before, the source is taken to be switched on
adiabatically from $t=-\infty$ and 
switched off at $t=0$ to provide initial conditions
\begin{eqnarray}
&&\psi_{\vec{k}}(t=0)=\psi_{\vec{k}}(0)~,\quad\dot{\psi}_{\vec{k}}(t <
0)=0~. \label{iniconds} 
\end{eqnarray}
Defining $\sigma_{\vec{k}}(t-t')$ as
\begin{equation}
\frac{d}{dt'}\sigma_{\vec{k}}(t-t')=\Sigma_{\vec{k}}(t-t')
\label{sigma}
\end{equation}
and imposing that $ \eta_{r,\vec{k}}(t>0)=0 $, we obtain the equation of
motion for  
$t>0$
\begin{eqnarray}
&&\left[\left(i\gamma_0\frac{\partial}{\partial t}
- \vec{\gamma}\cdot\vec{k}-M\right)+
\delta_\psi\left(i\gamma_0\frac{\partial}{\partial t}
- \vec{\gamma}\cdot\vec{k}\right)
+ \sigma_{\vec{k}}(0) -\delta_M \right] \psi_{\vec{k}}(t) -
\int_0^t dt'\; \sigma_{\vec{k}}(t-t')\;
\dot{\psi}_{\vec{k}}(t')=0~\;.\label{eom}
\end{eqnarray}
The equation of motion (\ref{eom}) can be solved by Laplace transform
as befits an initial value problem.
The Laplace transformed equation of motion is given by
\begin{eqnarray}
&&\left[i\gamma_0 s
- \vec{\gamma}\cdot\vec{k}-M+ \delta_\psi 
\left(i \gamma_0 s -\vec{\gamma}\cdot\vec{k}\right)-\delta_M 
+\sigma_{\vec{k}}(0) - s \tilde{\sigma}_{\vec{k}}(s)
\right] \tilde{\psi}_{\vec{k}}(s)\nonumber\\
&& \quad\quad\quad= 
\left[i\gamma_0 + i\delta_\psi \gamma_0- 
\tilde{\sigma}_{\vec{k}}(s)\right]\psi_{\vec{k}}(0)~\; ,
\end{eqnarray}
where $\tilde{\psi}_{\vec{k}}(s)$ and $\tilde{\sigma}_{\vec{k}}(s)$ 
are the Laplace transforms of $\psi_{\vec{k}}(t)$ and
$\sigma_{\vec{k}}(t)$ respectively:
$$
\tilde{\psi}_{\vec{k}}(s)\equiv \int_0^{\infty} dt \; e^{-st}\;
\psi_{\vec{k}}(t) \quad , \quad \tilde{\sigma}_{\vec{k}}(s)\equiv
\int_0^{\infty} dt \; e^{-st}\; \sigma_{\vec{k}}(t)\; .
$$

%With eq.~(\ref{sigma}), $\sigma_{\vec{k}}(t-t')$ can be written as
%\begin{eqnarray}
%&&\sigma_{\vec{k}}(t-t')=i\gamma_0 \sigma^{(0)}_{\vec{k}}(t-t')+
%\sigma^{(1)}_{\vec{k}}(t-t')\,\vec{\gamma}\cdot\vec{k}+
%\sigma^{(2)}_{\vec{k}}(t-t')~.
%\end{eqnarray}

\subsection{Renormalization}

Before proceeding with the solution of the above equation, we address
the issue of the renormalization by analyzing the ultraviolet
divergences of the kernels.  
As usual the ultraviolet divergences are those of zero
temperature field theory, since the finite temperature distribution
functions are exponentially suppressed at large momenta. Therefore the
ultraviolet divergences are obtained by setting to zero the bosonic and
fermionic occupation numbers. 

With eq.~(\ref{sigma}), $\sigma_{\vec{k}}(t-t')$ can be written as
\begin{eqnarray}
&&\sigma_{\vec{k}}(t-t')=i\gamma_0 \sigma^{(0)}_{\vec{k}}(t-t')+
\vec{\gamma}\cdot\vec{k}~\sigma^{(1)}_{\vec{k}}(t-t')+
\sigma^{(2)}_{\vec{k}}(t-t')~.
\end{eqnarray}
A straightforward calculation leads to
\begin{eqnarray}
&\sigma^{(0)}_{\vec{k}}(0) = 0 ~, \; \; 
\sigma^{(1)}_{\vec{k}}(0)=
-\frac{y^2}{16\pi^2}\ln\left(\frac{\Lambda}{{\cal
K}}\right)+\mbox{finite} ~,\;\;
\sigma^{(2)}_{\vec{k}}(0) =  \frac{y^2 M}{8
\pi^2}\ln\left(\frac{\Lambda}{{\cal K}}\right)+\mbox{finite}~,&\nonumber\\
&\tilde{\sigma}^{(0)}_{\vec{k}}(s)= -
 \frac{y^2}{16\pi^2}\ln\left(\frac{\Lambda}{{\cal
K}}\right)+\mbox{finite} ~,\;\;
\tilde{\sigma}^{(1)}_{\vec{k}}(s)= \mbox{finite} ~,\;\; 
\tilde{\sigma}^{(2)}_{\vec{k}}(s)= \mbox{finite}~,&\nonumber
\end{eqnarray}
where $\tilde{\sigma}^{(0)}_{\vec{k}}(s)$, 
$\tilde{\sigma}^{(1)}_{\vec{k}}(s)$ and 
$\tilde{\sigma}^{(2)}_{\vec{k}}(s)$
are the Laplace transform of $\sigma^{(0)}_{\vec{k}}(t)$, 
$\sigma^{(1)}_{\vec{k}}(t)$ and $\sigma^{(2)}_{\vec{k}}(t)$ 
respectively, 
$\Lambda$ is an ultraviolet momentum cutoff
and ${\cal K}$ is an arbitrary renormalization scale.
Therefore the counterterms $\delta_\psi$ and $\delta_M$ are chosen to be
given by 
\begin{eqnarray}
&&\delta_\psi = -\frac{y^2}{16\pi^2}\ln\left(\frac{\Lambda}{{\cal
K}}\right)+\mbox{finite} ~,\quad
\delta_M = \frac{y^2 M}{8 \pi^2} \ln\left(\frac{\Lambda}{{\cal
K}}\right)+\mbox{finite}~, \label{counterterms}
\end{eqnarray}
and the respective kernels are rendered finite, i.e., 
\begin{eqnarray}
\tilde{\sigma}_{\vec{k}}(s)-i \delta_\psi\gamma_0 =
\tilde{\sigma}_{r,\vec{k}}(s)= \text{finite}~,\quad
\sigma_{\vec{k}}(0) - \delta_\psi \vec{\gamma}\cdot\vec{k}
-\delta_M=\sigma_{r,\vec{k}}(0)
= \text{finite}~. \label{renker}
\end{eqnarray}
The finite parts of the counterterms in eq.~(\ref{counterterms}) are 
fixed by prescribing a renormalization scheme.
There are {\em two} important choices of counterterms: 
i) determining the counterterms from an on-shell condition, 
including finite temperature effects, and
ii) determining the counterterms from a zero temperature on-shell condition. 
Obviously these choices only differ by finite quantities, however the
second choice  allows us to separate the dressing
effects of the medium from those in the vacuum. 
For example by choosing to renormalize with the zero temperature
counterterms on-shell, the pole in the  particle propagator will have
unit residue  at zero temperature; however in the medium, 
the residue at the finite temperature poles (or the position of the
resonances)  are {\em finite}, smaller than one and determined solely by the
properties of the medium.  Thus the formulation of the
initial value problem as presented here yields an unambiguous separation
of the vacuum and in-medium renormalization effects. 

Hence we obtain the renormalized effective Dirac equation in the medium
and its initial value problem 
\begin{eqnarray}
&&\left[i\gamma_0 s - \vec{\gamma}\cdot\vec{k}-M
+\tilde{\Sigma}_{r,\vec{k}}(s)\right] \tilde{\psi}_{\vec{k}}(s) = 
\left[i\gamma_0 - 
\tilde{\sigma}_{r,\vec{k}}(s)\right]\psi_{\vec{k}}(0)~,\label{reneom} \\
&&\tilde{\Sigma}_{r,\vec{k}}(s)=
\sigma_{r,\vec{k}}(0)-s\tilde{\sigma}_{r,\vec{k}}(s)~, \nonumber
\end{eqnarray}
with $\sigma_{r,\vec{k}}(0)$ and $\tilde{\sigma}_{r,\vec{k}}(s)$ the fully 
renormalized kernels (see eq.~(\ref{renker})), and
$\tilde{\Sigma}_{r,\vec{k}}(s)$  
the Laplace transform of  the  renormalized fermion self-energy which
can be written in its most general form as follows
\begin{equation}
\tilde{\Sigma}_{r,\vec{k}}(s)=
i\; \gamma_0\;s \;\tilde{\varepsilon}^{(0)}_{\vec{k}} (s)+
\vec{\gamma}\cdot \vec{k}~~\tilde{\varepsilon}^{(1)}_{\vec{k}}(s)+M \;
\tilde{\varepsilon}^{(2)}_{\vec{k}}(s)~.\label{epsilons}
\end{equation}

The solution to eq.~(\ref{reneom}) is 
\begin{equation}
\tilde{\psi}_{\vec{k}}(s)= \frac{1}{s}\left[ 1 + S(s,\vec{k}) 
\left(\vec{\gamma}\cdot\vec{k}+M -\tilde{\Sigma}_{r,\vec{k}}(0)\right)\right]
\psi_{\vec k}(0)~,
\label{laplakern}
\end{equation}
where
\begin{eqnarray}
S(s,\vec{k})& = & \left[i\gamma_0 s - \vec{\gamma}\cdot\vec{k}-M
+\tilde{\Sigma}_{r,\vec{k}}(s)\right]^{-1} \nonumber \\
& = & 
\frac{i\gamma_0 s(1+\tilde{\varepsilon}^{(0)}_{\vec{k}}(s)) - 
\vec{\gamma}\cdot\vec{k}(1-\tilde{\varepsilon}^{(1)}_{\vec{k}}(s))
+M(1-\tilde{\varepsilon}^{(2)}_{\vec{k}}(s))}{
-s^2(1+\tilde{\varepsilon}^{(0)}_{\vec{k}}(s))^2
-k^2(1-\tilde{\varepsilon}^{(1)}_{\vec{k}}(s))^2
-M^2(1-\tilde{\varepsilon}^{(2)}_{\vec{k}}(s))^2}\label{fermpropofs}
\end{eqnarray}
is the fermion propagator in terms of the Laplace variable $s$. The square of the denominator of 
(\ref{fermpropofs}) is recognized as 
$$ 
\det \left[i\gamma_0 s -\vec{\gamma}\cdot\vec{k}-M 
+\tilde\Sigma_{r,\vec{k}}(s)\right]
$$ 
The real-time evolution of $\psi_{\vec{k}}(t)$ is obtained by performing
the inverse Laplace transform along the Bromwich contour in the
complex $s$-plane parallel to the imaginary axis and to the right of all 
singularities of $\tilde{\psi}_{\vec{k}}(s)$. Therefore to obtain the real 
time evolution we must first understand the singularities of the Laplace
transform in the complex $s$-plane.

\section{Structure of the self-energy and damping processes}
To one loop order, the Laplace transform of the components $\tilde{\varepsilon}^{(i)}_{\vec{k}}(s)$ of the
fermion self-energy
$\tilde{\Sigma}_{\vec k}(s)$ (see eq.~(\ref{epsilons})) can be written as  dispersion
integrals in terms of  spectral densities $\rho^{(i)}_{\vec{k}}(k_0)$
\begin{equation}
\left\{ \begin{array}{c}
\tilde{\varepsilon}^{(0)}_{\vec{k}}(s)  \\
\tilde{\varepsilon}^{(1)}_{\vec{k}}(s)  \\
\tilde{\varepsilon}^{(2)}_{\vec{k}}(s)  
\end{array} \right\} = 
 \int dk_0 \frac{1}{s^2+k_0^2} \left\{\begin{array}{c}
\rho^{(0)}_{\vec{k}}(k_0)  \\
k_0\; \rho^{(1)}_{\vec{k}}(k_0)  \\
k_0\; \rho^{(2)}_{\vec{k}}(k_0)  
\end{array} 
\right\}+
\left\{\begin{array}{c}
\delta_{\psi}  \\
-\delta_{\psi} \\
-\frac{\delta_M}{M}  
\end{array} 
\right\}~,\nonumber
%\label{dispersion}
\end{equation}
with the one-loop  spectral densities given  by the expressions
\begin{eqnarray}
\rho^{(0)}_{\vec{k}}(k_0)&=& y^2 \int \frac{d^3 q}{(2\pi)^3
}\frac{\bar{\omega}_{q}}{
2 \omega_{k+q} \bar{\omega}_{q}} \times\nonumber\\
&&\left[\delta(k_0-\omega_{k+q}-\bar{\omega}_{q})
(1+n_{k+q}-\bar{n}_{q})+ \delta(k_0-\omega_{k+q}+\bar{\omega}_{q})
(n_{k+q}+\bar{n}_{q}) \right]~,
%\label{disprel}
\nonumber\\ \rho^{(1)}_{\vec{k}}(k_0)&=& y^2 \int \frac{d^3 q}{(2\pi)^3
} \frac{1}{2 \omega_{k+q} \bar{\omega}_{q}}\frac{\vec k \cdot \vec
q}{k^2}\times\nonumber\\ 
&&\left[\delta(k_0-\omega_{k+q}-\bar{\omega}_{q})
(1+n_{k+q}-\bar{n}_{q})-\delta(k_0-\omega_{k+q}+\bar{\omega}_{q})
(n_{k+q}+\bar{n}_{q})\right]~, \nonumber \\
\rho^{(2)}_{\vec{k}}(k_0)&=& y^2 \int \frac{d^3 q}{(2\pi)^3
} \frac{1}{
2 \omega_{k+q} \bar{\omega}_{q}}  \times \nonumber\\
&&\left[\delta(k_0-\omega_{k+q}-\bar{\omega}_{q})
(1+n_{k+q}-\bar{n}_{q})-\delta(k_0-\omega_{k+q}+\bar{\omega}_{q})
(n_{k+q}+\bar{n}_{q})\right]~.\label{rhos}
\end{eqnarray}

The analytic continuation of the self-energy and its components 
$\tilde{\varepsilon}^{(i)}(s)$ in the complex $s$-plane are given by
\begin{eqnarray}
&& \tilde{\Sigma}_{r,\vec k}(s=-i\omega \pm 0^+) =
\Sigma_{R,k}(\omega)\pm i \;  
\Sigma_{I,k}(\omega)~, \nonumber \\
&&\tilde{\varepsilon}^{(i)}_{\vec{k}}(s=-i\omega\pm 0^+) 
= \varepsilon^{(i)}_{R,\vec{k}}(\omega)\pm
i\; \varepsilon^{(i)}_{I,\vec{k}}(\omega)~\; , 
\label{realandimaginary}
\end{eqnarray}
where the real parts are even functions of $\omega$ and given by
\begin{equation}
\left\{ \begin{array}{c}
\varepsilon^{(0)}_{R,\vec{k}}(\omega)  \\
\varepsilon^{(1)}_{R,\vec{k}}(\omega)  \\
\varepsilon^{(2)}_{R,\vec{k}}(\omega)  
\end{array} \right\} = 
 \int dk_0 {\cal P}\left(\frac{1}{k_0^2-\omega^2}\right) \left\{\begin{array}{c}
\rho^{(0)}_{\vec{k}}(k_0)  \\
k_0\rho^{(1)}_{\vec{k}}(k_0)  \\
k_0\rho^{(2)}_{\vec{k}}(k_0)  
\end{array} 
\right\} +
\left\{\begin{array}{c}
\delta_{\psi}  \\
-\delta_{\psi} \\
-\frac{\delta_M}{M}  
\end{array} 
\right\}~,
\label{realpart}
\end{equation}
and the imaginary parts  are odd functions of $\omega$ and given by
\begin{eqnarray}
\varepsilon^{(0)}_{I,\vec{k}}(\omega) & = &
\frac{\pi}{2|\omega|}\mbox{sgn}(\omega)\left[ 
\rho^{(0)}_{\vec{k}}(|\omega|)+\rho^{(0)}_{\vec{k}}(-|\omega|)\right]~,
\nonumber \\ 
\varepsilon^{(1)}_{I,\vec{k}}(\omega) & = &
\frac{\pi}{2}\mbox{sgn}(\omega)\left[ 
\rho^{(1)}_{\vec{k}}(|\omega|)-
\rho^{(1)}_{\vec{k}}(-|\omega|)\right]~, \nonumber   \\ 
\varepsilon^{(2)}_{I,\vec{k}}(\omega)  & = &
\frac{\pi}{2}\mbox{sgn}(\omega)\left[ 
\rho^{(2)}_{\vec{k}}(|\omega|)- \rho^{(2)}_{\vec{k}}(-|\omega|)\right]~.
\label{imagipart}  
\end{eqnarray}

The denominator of the analytically continued fermion propagator
eq.~(\ref{fermpropofs}) can be written in the compact form
$$\omega^2 - \bar{\omega}^2_k+\Pi(\omega,\vec{k})$$
with $\bar{\omega}_k= \sqrt{\vec{k}^2+M^2}$ and 
\begin{eqnarray}
\Pi(\omega,\vec{k}) &=&2\left[\omega^2 \varepsilon^{(0)}_{\vec{k}}(\omega)+k^2\varepsilon^{(1)}_{\vec{k}}(\omega)+M^2
\varepsilon^{(2)}_{\vec{k}}(\omega)\right] \nonumber\\
&+&\omega^2 \; [\varepsilon^{(0)}_{\vec{k}}(\omega)]^2
-k^2 \;[\varepsilon^{(1)}_{\vec{k}}(\omega)]^2- 
M^2 \; [\varepsilon^{(2)}_{\vec{k}}(\omega)]^2 ~. \label{totalpi}
\end{eqnarray}

We recognize that the {\em lowest order term} of this effective  self-energy  
can be written in the familiar form~\cite{weldon2,kapusta,lebellac}
\begin{equation}
\Pi(\omega,\vec{k}) = 2\left[\omega^2 \varepsilon^{(0)}_{\vec{k}}(\omega)+k^2\varepsilon^{(1)}_{\vec{k}}(\omega)+M^2
\varepsilon^{(2)}_{\vec{k}}(\omega)\right]=  
\frac{1}{2}\text{Tr}[
(\not\!k+M)\tilde{\Sigma}_{\vec{k}}(s=-i\omega+0^+)]~,\label{Pi}
\end{equation}
but certainly {\em not the higher order terms}. 

The imaginary part of $\Pi$ on-shell will be identified with the
damping rate (see below).  The expression given 
by eq.~(\ref{Pi}) leads to  the familiar form of the damping rate, but
we point out that eq.~(\ref{Pi}) is a {\em lowest order} result. 
The full imaginary part must be obtained from the full function
$\Pi(\omega,\vec{k})$ and the generalization to {\em all orders} will
be given in  a later section below.

For fixed $M$ and $m$, 
the $\delta$-function constraints in the spectral densities 
$\rho^{(i)}_{\vec{k}}(\omega)$ can only be satisfied for certain
ranges of $\omega$. 
Since the imaginary parts are
odd functions of $\omega$ we only consider the case of positive $\omega$. 
The $\delta(|\omega|-\omega_{k+q}-\bar{\omega}_q)$ has support 
only for $|\omega| > \sqrt{k^2+(m+M)^2}$ and  corresponds to the normal  
two-particle cuts that are present at zero temperature corresponding to the 
process $\psi \rightarrow \phi + \psi$. These cuts (both
for positive and negative $\omega$) 
do not give a contribution to the imaginary part on-shell for the
fermion.

The terms proportional to $ n_{k+q}+\bar{n}_q $ give the following
contribution to the lowest order effective self-energy:

\begin{eqnarray}
\Pi_I(\omega,\vec{k}) &=& \pi y^2 \text{sgn}(\omega) \int \frac{d^3q}{(2\pi)^3} 
\frac{n_{k+q}+\bar{n}_q }{2\bar{\omega}_q \omega_{k+q}}\left[
\left(|\omega| \bar{\omega}_q-\vec k \cdot \vec q -M^2\right)\delta
\left(|\omega|-\omega_{k+q}+\bar{\omega}_q\right) \right. \nonumber \\
&&\quad\quad\quad\quad\quad+\left. \left(|\omega|\bar{\omega}_q+\vec k \cdot \vec q +M^2\right)
\delta\left(|\omega|+\omega_{k+q}-\bar{\omega}_q\right)\right]~.
\label{Piimag}
\end{eqnarray}

\noindent The first delta function determines a cut in the region 
$0<|\omega|< \sqrt{k^2+(m-M)^2}$ and originates in the physical process 
$\phi \rightarrow \psi + \bar{\psi}$ whereas the
second delta function determines a cut in the region $0<|\omega|<k$ 
and originates in the process
$\phi + \psi \rightarrow \psi$. 
Whereas the first cut originates in the process of decay of the scalar
into fermion pairs,  
the second cut for ( $\omega^2<k^2$) is associated with  Landau damping. 
Both delta functions restrict the range of the integration
variable $q$ (see below). 
For $m>2M$ the scalar can decay on-shell into a fermion pair, 
and in this case the fermion pole is embedded in the cut 
$0<|\omega|< \sqrt{k^2+(m-M)^2}$ becoming a quasiparticle pole and only the 
first cut contributes to the quasiparticle width. 
This is a remarkable result, {\em the fermions acquire a width through the
induced decay of the scalar in the medium}. This process only occurs in
the medium (obviously vanishing at $T=0$) and its origin is very different
from either collisional broadening or Landau damping. 
A complementary interpretation of the origin of this
process as a medium induced decay of the scalars into fermions and the 
resulting quasiparticle width for the fermion excitation will 
be highlighted in section V within the kinetic approach to relaxation. 

The width is obtained to lowest order from
$\Pi_I(\omega_k,\vec{k})$, and the on-shell delta function is recognized 
as the energy conservation condition for the
{\em decay} $\phi \rightarrow \psi +\bar{\psi}$ of the heavy scalar on-shell. 
To lowest order we find the following expression for
$\Pi_I(\bar{\omega}_k,\vec{k})$  
for arbitrary scalar and fermion masses with $m>2M$ and arbitrary fermion
momentum and temperature:

\begin{eqnarray}
\Pi_I(\bar{\omega}_k,\vec{k})&=&\pi y^2\int \frac{d^3q}{ (2\pi)^3 }
\frac{\bar{\omega}_k \bar{\omega}_q -
\vec{k}\cdot\vec{q}-M^2}{2\bar{\omega}_q \omega_{k+q}}  \;
(n_{k+q}+\bar{n}_q)\;
\delta(\bar{\omega}_{k}+\bar{\omega}_{q}-\omega_{k+q})~,\nonumber\\
&=&\frac{y^2 m^2 T}{16 \pi k }\left(1-\frac{4M^2}{m^2}\right)
\ln\left.\left[\frac{1-e^{-\beta(\bar{\omega}_{q}+\bar{\omega}_{k})}}{1+
e^{-\beta\bar{\omega}_{q}}}\right]\right |
^{\bar{\omega}_{q^{\ast}_{2}}}_{\bar{\omega}_{q^{\ast}_{1}}}~,
\label{dampingrate}
\end{eqnarray}
where $q^{\ast}_{1}$ and $q^{\ast}_{2}$ are given by
\begin{eqnarray}
q^{\ast}_{1}&=&
\frac{m^2}{2 M^2} \left| k\left(1-\frac{2M^2}{m^2}\right)-
\sqrt{(k^2+M^2)\left(1-\frac{4M^2}{m^2}\right)}\right|~,\nonumber \\
q^{\ast}_{2}&=& \frac{m^2}{2 M^2} \left[
k\left(1-\frac{2M^2}{m^2}\right)+
\sqrt{(k^2+M^2)\left(1-\frac{4M^2}{m^2}\right)}\,\right]\label{q2star}~,
\end{eqnarray} 
with $|\vec{q}|\in (q^{\ast}_{1},q^{\ast}_{2})$ being the support of
$\delta(\bar{\omega}_k-\omega_{k+q}+\bar{\omega}_{q})$.

\section{Real time evolution}

The real time evolution is obtained by performing the inverse Laplace transform 
as explained in section II. This requires analyzing the singularities of 
$\tilde{\psi}_{\vec k}(s)$ given by eq.~(\ref{fermpropofs}) in the complex $s$-plane.
It is straightforward to see that the putative pole at $s=0$ has vanishing residue, 
therefore the singularities are those arising from the inverse
fermion propagator $S(s,\vec{k})$. If the fermionic pole is away from 
the multiparticle cuts, the singularities are: 
i) the isolated fermion poles at $s= - i \omega_{\text{p}}$ with 
$\omega_{\text{p}}$ the position of the isolated (complex) poles
(corresponding to stable fermionic excitations), and 
ii) the multiparticle cuts along the imaginary axis $s= -i\omega$ 
for $0<|\omega|< \sqrt{k^2+(m-M)^2}$ and $|\omega|>\sqrt{k^2+(m+M)^2}$. 

The Laplace transform is performed by deforming the contour, circling the
isolated poles and wrapping around the cuts. When the scalar 
particle can decay into fermion pairs, i.e., $ m>2M $, the
fermion pole is embedded in the lower cut and we must find out if 
it becomes a complex pole in the physical sheet (the domain of integration) 
or moves off the physical sheet. 

For $ m < 2M $ the fermion pole at $ \omega = \omega_{\text{p}} $ is
real and the fermion mean field oscillates for late times with constant
amplitude and frequency $ \omega_{\text{p}} $.

\subsection{$m>2M$: complex poles or resonances?} 

When $m>2M$ the fermion pole is embedded in the cut 
$0<|\omega|<\sqrt{k^2+(m-M)^2}$ and the pole becomes complex. 
The position of the complex poles are obtained from 
the zeros of $\omega^2 -\bar{\omega}^2_k +\Pi(\omega,\vec{k})$ in the analytically
continued fermion propagator for $\omega=\omega_{\text{p}}(k) -i\Gamma_k$ with 
$\omega_{\text{p}}$ being the real part of the complex
pole. For $\Gamma_k \ll \omega_{\text{p}}$ (narrow width approximation) 
and with the expressions for the discontinuities in the physical 
sheet given by eq.~(\ref{realandimaginary}), the equation that determines 
the position of the complex pole is given by the solution of the following equation
\begin{equation}
(\omega_{\text{p}} -i\;\Gamma_k)^2 -\bar{\omega}^2_k +
\Pi_R(\omega_{\text{p}},\vec{k})-i \; 
\mbox{sgn}(\Gamma_k)\;\Pi_I(\omega_{\text{p}},\vec{k})=0~, \label{complexpole}
\end{equation} 
where we have used the narrow width approximation. 
To lowest order, the real and imaginary parts of this equation become
\begin{eqnarray}
&&\omega^2_{\text{p}}
-\bar{\omega}^2_k+\Pi_R(\omega_{\text{p}},\vec{k}) = 0~,
\label{realpole}\\ 
&& \Gamma_k =
-\mbox{sgn}(\Gamma_k)\;
\frac{\Pi_I(\omega_{\text{p}},\vec{k})}{2\;\omega_{\text{p}}}~,
\label{imagipole}
\end{eqnarray}
and the lowest order solution of eq.~(\ref{realpole}) is given by
\begin{equation}
\omega_{\text{p}} = \pm\left[\bar{\omega}_k -
\frac{\Pi_R(\bar{\omega}_k,\vec{k})}{2\bar{\omega}_k} \right] 
=\pm\left[ \bar{\omega}_{k} - \frac{1}{4\bar{\omega}_{k}}\text{Tr}[
(\not\!k+M)\Sigma_{R,\vec{k}}(\bar{\omega}_k)]\right]~. \label{lowestorderpole}
\end{equation}
The solution for the imaginary part is obtained by replacing 
$\omega_{\text{p}} = \pm\bar{\omega}_k$ in $\Pi_I(\omega_{\text{p}},\vec{k})$ 
to this order.  
However, the equation for the imaginary part (\ref{imagipole}) 
{\em does not have a solution} because $\Pi_I(\omega,\vec{k})$ is an
odd function of $\omega$ and $\Pi_I(\bar{\omega}_k,\vec{k})>0$. 
Therefore, there is {\em no} complex pole in the physical sheet. This
is a fairly  well-known (but seldom noticed) result: if the imaginary
part of the self-energy on shell  is positive there is no complex
pole solution in the physical sheet, the pole has moved off into the
unphysical (second) sheet.  

In the case that  $ \Pi_I(\bar{\omega}_k,\vec{k}) $ is negative, 
complex poles  appear in the physical sheet,  
but in such  case there are {\em two poles}  with both signs for
$ \Gamma_k $, i.e.,  
one corresponds to a growing exponential in time and the other a decaying
exponential, this is the signal of an instability, not of damping. 
However since we confirm that $ \Pi_I(\bar{\omega}_k,\vec{k}) $ is positive in the
case under consideration, the complex poles are in the unphysical sheet, describing
a resonance.  

\subsection{Scalar decay implies fermion damping}

Since we have determined that there are {\em no} complex poles in the
physical sheet in the $s$-plane, i.e., the integration region, the only singularities are the
cut discontinuities along  the following segments of the imaginary axis: 
$$
s=\left[i\sqrt{k^2+(m+M)^2},i\infty\right] \; , \;
s=\left[-i\sqrt{k^2+(m+M)^2},-i\infty\right]
$$ 
and
$$
s=\left[-i\sqrt{k^2+(m-M)^2} , i\sqrt{k^2+(m-M)^2}\right]\; .
$$ 
The contour of integration (Bromwich contour) 
is deformed to wrap around these cuts. In the narrow width approximation
and consistent with perturbation theory, the discontinuities in the 
self-energy are perturbatively small; and since either the real pole or
the resonance is below the two particle cut, the contribution from this cut 
is always perturbatively small. On the other hand, in the relevant case of 
$m>2M$ with a quasiparticle resonance, the contribution from the cut
$$
s=\left[-i\sqrt{k^2+(m-M)^2},i\sqrt{k^2+(m-M)^2}\right]
$$ 
becomes the dominant one. It is convenient to write the product in
eq.~(\ref{laplakern}) in the simplified form 
\begin{equation}
S(s,\vec{k}) 
\left(\vec{\gamma}\cdot\vec{k}+M -\tilde{\Sigma}_{r,\vec{k}}(0)\right)
= \frac{\tilde{{\cal N}}(s,k)}{-s^2 -\bar{\omega}^2_k +\Pi(s,k)}~,
\label{numerator}  
\end{equation}
which defines $\tilde{{\cal N}}(s,k)$, and to change variables to
$s=-i\omega \pm 0^+$ on  both sides of the cut, 
leading to the following contribution from the thermal cut to the real
time evolution  
\begin{eqnarray}
\psi_{\vec k}(t) = -\frac{1}{\pi} \int_{\omega^-}^{\omega^+} 
\frac{d\omega}{\omega}e^{-i\omega t} && \left\{ 
\frac{{\cal N}_R(\omega,\vec{k}) \;
\Pi_I(\omega,\vec{k})}{[\omega^2-\bar{\omega}^2_k+ 
\Pi_R(\omega,\vec{k})]^2+\Pi_I(\omega,\vec{k})^2} \right. \nonumber \\
&&-\left. \frac{{\cal N}_I(\omega,\vec{k})\;
[\omega^2-\bar{\omega}^2_k+\Pi_R(\omega,\vec{k})] }
{[\omega^2-\bar{\omega}^2_k+\Pi_R(\omega,\vec{k})]^2+\Pi_I(\omega,\vec{k})^2}
\right\}  
\psi_{\vec k}(0)~,
\end{eqnarray}
where ${\cal N}_{R,I}(\omega,\vec{k})$ are obtained by replacing the 
$\varepsilon^{(i)}_{\vec{k}}(\omega)$ by their real or imaginary parts and 
$\omega^{\pm}= \pm \sqrt{k^2+(m-M)^2}$. The term proportional to 
${\cal N}_R(\omega,\vec{k})$ features a typical Breit-Wigner resonance
shape near the real part of the complex pole at 
$\omega^2_{\text{p}}-\bar{\omega}^2_k+\Pi_R(\omega_{\text{p}},\vec{k})=0$ 
since, for $m>2M$, the imaginary part of the
self energy at this value of $\omega$ (perturbatively close to $\pm\omega_k$) 
is non-vanishing. On the other hand, the term proportional to 
${\cal N}_I(\omega,\vec{k})$ 
is a representation of the principal part in the limit of small 
$\Pi_I(\omega,\vec{k})$ and is therefore subleading. The  sharply
peaked resonances at  
$\omega = \pm|\omega_{\text{p}}| \approx \pm \bar{\omega}_k$ dominate the
spectral density and give the largest contribution to the real time 
evolution. In the limit of a narrow resonance, the $\omega$ integral is
performed by taking the integration limits to infinity and 
approximating near the resonances at $\omega = \pm |\omega_{\text{p}}|$,
\begin{eqnarray}
&&\frac{1}{\omega}
\frac{\Pi_I(\omega,\vec{k})}{[\omega^2-\bar{\omega}^2_k+
\Pi_R(\omega,\vec{k})]^2+ \Pi_I(\omega,\vec{k})^2} \approx  
\frac{Z_k}{2\omega^2_{\text{p}}} \frac{\Gamma_k}{(\omega\mp
|\omega_{\text{p}}|)^2  + \Gamma^2_k}~, \nonumber \\
&& Z_k= \left[1+\left.\frac{\partial \Pi_R(\omega,\vec{k})}{\partial
\omega^2}\right|_{\omega_{\text{p}}} \right]^{-1}~, \; \; \; 
\Gamma_k = \frac{Z_k\Pi_I(\omega_{\text{p}},\vec{k})}{2
\omega_{\text{p}}}~. \label{breit} 
\end{eqnarray} 
The wave-function renormalization constant $Z_k$ 
is {\em finite} since the self-energy has been rendered finite
by an appropriate choice of counterterms. 
Only when the counterterms are chosen to provide a subtraction of the
self-energy  
at the position of the resonance $\omega_{\text{p}}$ will result in $Z_k=1$. On
the other hand,   
if the counterterms are chosen to renormalize the theory on-shell at {\em
zero temperature}, then $Z_k$ describes the dressing of the medium and
is smaller than one. 

The integration in the variable $\omega$ is performed under these
approximations (justified for narrow width) leading to the real time evolution
\begin{equation}
\psi_{\vec k}(t) \approx \frac{Z_k}{2\omega^2_{\text{p}}} \left[
{\cal N}_R(|\omega_{\text{p}}|,\vec{k})e^{-i|\omega_{\text{p}}| t}+
{\cal N}_R(-|\omega_{\text{p}}|,\vec{k})e^{i|\omega_{\text{p}}|
t}\right]e^{-\Gamma_k t}\; \psi_{\vec k}(0)~. \label{finalrealtime} 
\end{equation}
Gathering the results for the imaginary part of the effective self-energy in 
eq.~(\ref{dampingrate}), we find the real-time damping rate for the
fermionic mean fields to one loop order for arbitrary values of the
scalar and fermion masses (with $m>2M$), temperature and fermion
momentum to be given by 
\begin{eqnarray}
\Gamma_k&=&\pi y^2\int \frac{d^3q}{ (2\pi)^3 }
\frac{\bar{\omega}_k \bar{\omega}_q - 
\vec{k}\cdot\vec{q}-M^2}{2\bar{\omega}_k 2\bar{\omega}_q \omega_{k+q}} \;
(n_{k+q}+\bar{n}_q)\;
\delta(\bar{\omega}_{k}+\bar{\omega}_{q}-\omega_{k+q})\nonumber\\
&=&\frac{y^2 m^2 T}{32 \pi k \bar{\omega}_k }\left(1-\frac{4M^2}{m^2}\right)
\ln\left.\left[\frac{1-e^{-\beta(\bar{\omega}_{q}+\bar{\omega}_{k})}}{1
+e^{-\beta\bar{\omega}_{q}}}\right]\right |
^{\bar{\omega}_{q^{\ast}_{2}}}_{\bar{\omega}_{q^{\ast}_{1}}}~,
\label{dampingratefin}
\end{eqnarray}
where $q^{\ast}_{1}$ and $q^{\ast}_{2}$ are given by
eqs.~(\ref{q2star}).  

Figures~(1-3) display the behavior of $\Gamma_k$ for several ranges of
the parameters. We have chosen a wide range of parameters for the
ratios of the scalar  to fermion masses ($m/M$) and the
temperature to fermion mass ($T/M$) to illustrate in detail the
important differences.  The damping rate features a
strong peak as a function of the ratio $k/M$. This peak is at very
small momentum when the ratio of scalar to fermion
mass is not much larger than 2, but moves to larger values of the
fermion momentum when this ratio is very large. Figure~1 
displays this feature in an extreme case ($m/M=800$) to highlight this
behavior. The height of the peak is a monotonically increasing
function of temperature as expected.  

This is one of the important results of this work: the
{\em induced} decay of the heavy scalar into fermion pairs results in
a {\em damping} of the amplitude of fermionic excitations. 

\subsection{Resonance wave functions and all-order expression for the
damping rate} 

Since there is no complex pole solution in the physical sheet, 
there are no solutions of the effective in-medium 
Dirac equation for $m>2M$. 
However, we can define the spinor wave function of the 
{\em resonance} by considering the
solutions of the in-medium Dirac equation with only the {\em real part} of the 
self-energy~\cite{dolivo1,dolivodamp1} at the the value $\omega
=\omega_{\text{p}}$.  
Since the form given by eq.~(\ref{epsilons}) for the self-energy 
is general and not restricted to perturbation theory, 
our analysis below is valid to {\em all orders}.

Using the expression for the self-energy given by eq.~(\ref{epsilons}) 
for $s=-i\omega_{\text{p}}$ and
the real part of the coefficients $\varepsilon^{(i)}_{\vec
k}(\omega_{\text{p}})$  
given by eqs.~(\ref{realandimaginary}, \ref{realpart}), we now introduce  
the following variables
\begin{equation}
{\cal W}_{\text{p}} = \omega_{\text{p}}\;[1+\varepsilon^{(0)}_{R,\vec
k}(\omega_{\text{p}})]~, 
\quad
\vec{{\cal K}} = \vec k\; [1-\varepsilon^{(1)}_{R,\vec
k}(\omega_{\text{p}})]~,\quad 
{\cal M} = M\; [1-\varepsilon^{(2)}_{R,\vec k}(\omega_{\text{p}})]~,
\label{newvariables} 
\end{equation}
then the resonance wave functions $\Psi_{\vec k}(\omega_{\text{p}})$ obey the
following 
effective in-medium Dirac equation
\begin{equation}
\left[\gamma_0 {\cal W}_{\text{p}} - \vec{\gamma}\cdot \vec{{\cal K}}  
- {\cal M}\right]\Psi_{\vec k}(\omega_{\text{p}})=0~. \label{diracsolution}
\end{equation}
Using the property $\varepsilon^{(i)}_R(\omega) = \varepsilon^{(i)}_R(-\omega)$ 
(see eq.~(\ref{realpart})), we define the two particle solutions as 
$U_{s;\vec k}=\Psi_{\vec k}(\omega_{\text{p}})$ for $\omega_{\text{p}}=|\omega_{\text{p}}|$ and the
two antiparticle solutions as $V_{s;\vec k}= \Psi_{-\vec
k}(\omega_{\text{p}})$ for  
$\omega_{\text{p}}= - |\omega_{\text{p}}|$. These in-medium solutions satisfy 
\begin{eqnarray} 
&&({\not\!{\cal K}}-{\cal M})U_{s;\vec k} =   0 \quad \text{and}\quad
({\not\!{\cal K}}+{\cal M})V_{s;\vec k} =   0~,\label{parantipar} 
\end{eqnarray}
with ${\not\!{\cal K}}  = \gamma_0 |{\cal
W}_{\text{p}}|-\vec{\gamma}\cdot \vec{{\cal K}}$.  

These Dirac spinors can be obtained from the usual free particle solutions
by the replacement 
$\bar{\omega}_k \rightarrow |{\cal W}_{\text{p}}|$, $\vec k
\rightarrow \vec{{\cal K}}$  
and $M \rightarrow {\cal M}$ 
given by  eq.~(\ref{newvariables}).  
Combining eqs. (\ref{epsilons}, \ref{realandimaginary}, \ref{totalpi}) and
the definition of the width given by (\ref{breit}), we
obtain the following expression for the width of the resonance 
{\em to all orders in perturbation theory\/}
\begin{equation}
\Gamma_k = \frac{Z_k}{4|\omega_{\text{p}}|} 
\mbox{Tr}\left[({\not\!{\cal K}}+{\cal
M})\Sigma_I(|\omega_{\text{p}}|)\right]=  
\frac{Z_k {\cal M}}{2|\omega_{\text{p}}|}\sum_{s=\pm}\left[ \bar{U}_{s;\vec k} 
\Sigma_I(|\omega_{\text{p}}|)U_{s;\vec k} \right]~, \label{exactwidth}
\end{equation}
where we have alternatively written the expression for the width in
terms of the  
{\em exact} resonance spinors in the medium. This result  
confirms those found in reference~\cite{dolivo1,dolivodamp1} 
and leads to the often quoted expression for the width~\cite{weldon2,lebellac}
in lowest order.  

\section{Kinetics of fermion relaxation}

To clarify and confirm independently the result of the previous
section of induced decay
of the scalar leading to a quasiparticle width of the fermionic excitations, 
we now provide an analysis of the relaxation of the distribution function for 
the fermions via a kinetic Boltzmann equation. This analysis will also
provide a  
firm relationship between the damping rate and the interaction rate in
real time  
in the relaxation time approximation (linearization near equilibrium). 

Let us denote  
the distribution function for scalars of momentum $\vec{k}$ at time $t$ by $N_{\vec{k}}(t)$
and that for fermions of momentum $\vec{k}$ 
and spin $s$ by $\bar{N}_{\vec{k},s}(t)$. 
In the kinetic approach, the derivative of this distribution function 
with respect to time is obtained from a Boltzmann equation. 
Since for a fixed spin component the matrix elements
for the transition probabilities are rather cumbersome, we define the
spin-averaged fermion distribution function as $ \bar{N}_{\vec{k}}(t) = \frac{1}{2}
\sum_{s} \bar{N}_{\vec{k},s}(t) $. 

Two processes are responsible for the change in the 
fermion populations in a fermion-scalar plasma: i)
$\phi\rightarrow \psi+\bar\psi$ (creation) which provides the `gain' 
term in the balance equation, and ii) $\psi+\bar\psi \rightarrow \phi$
(annihilation) which provides the `loss' term. Using the standard approach 
to obtain kinetic Bolzmann rate equations,
we find that the spin-averaged rates for creation and annihilation are given by 
\begin{eqnarray}
\left. \frac{d}{dt}\bar{N}_{\vec{k}}(t) \right|_{\text{gain}}
& = &\pi y^2\int \frac{d^3q}{(2 \pi)^3 2\bar{\omega}_q} 
\frac{\bar{\omega}_k \bar{\omega}_q - \vec{k}\cdot\vec{q}-M^2}
{\bar{\omega}_k \omega_{k+q}} \times \nonumber\\
&&N_{\vec{k}+\vec{q}}(t)[1-\bar{N}_{\vec{k}}(t)]
[1-\bar{N}_{\vec{q}}(t)]
\delta(\bar{\omega}_{k}+\bar{\omega}_{q}-\omega_{k+q})~,\nonumber\\
\left. \frac{d}{dt}\bar{N}_{\vec{k}}(t) \right|_{\text{loss}} &=& 
\pi y^2\int \frac{d^3q}{(2 \pi)^3 2\bar{\omega}_q} 
\frac{\bar{\omega}_k \bar{\omega}_q - \vec{k}\cdot\vec{q}-M^2}
{\bar{\omega}_k \omega_{k+q}} \times \nonumber\\
&&[1+N_{\vec{k}+\vec{q}}(t)]\bar{N}_{\vec{k}}(t)
\bar{N}_{\vec{q}}(t)
\delta(\bar{\omega}_{k}+\bar{\omega}_{q}-\omega_{k+q})~,\nonumber
\end{eqnarray}
respectively. The spin-averaged net rate is simply
\begin{eqnarray}
\frac{d}{dt}\bar{N}_{\vec{k}}(t) &=&\left.
\frac{d}{dt}\bar{N}_{\vec{k}}(t) \right|_{\text{gain}}-
\left.\frac{d}{dt}\bar{N}_{\vec{k}}(t) \right|_{\text{loss}}~.
%&=& \pi y^2\int \frac{d^3q}{(2 \pi)^3 2\bar{\omega}_q} 
%\frac{(\bar{\omega}_k \bar{\omega}_q - \vec{k}\cdot\vec{q}-M^2)}
%{\bar{\omega}_k \omega_{k+q}}
%\delta(\bar{\omega}_{k}+\bar{\omega}_{q}-\omega_{k+q})  
%\times \nonumber\\
%&&\Bigl\{N_{\vec{k}+\vec{q}}(t)[1-\bar{N}_{\vec{k}}(t)][1-\bar{N}_{\vec{q}}(t)
%]-
%[1+N_{\vec{k}+\vec{q}}(t)]\bar{N}_{\vec{k}}(t) \bar{N}_{\vec{q}}(t) \Bigr\}~,
%\frac{y^2 m^2}{16 \pi k
%\bar{\omega}_k}\left(1-\frac{4M^2}{m^2}\right) 
%\int^{\bar{\omega}_{q^{\ast}_{2}}}_{\bar{\omega}_{q^{\ast}_{1}}}
%d\bar{\omega}_{q}
%\left[n_{\bar{\omega}_{q}+\bar{\omega}_{k}}(1-\bar{n}_k)(1-\bar{n}_q)-
%(1+n_{\bar{\omega}_{q}+\bar{\omega}_{k}})\bar{n}_k\bar{n}_q\right]~,
\end{eqnarray}

Let us now consider that all of the modes but the fermionic mode with
wavevector  
$\vec{k}$ in the fermion-scalar plasma are in thermal
equilibrium, while the population for the fermionic $\vec{k}$-mode has a small 
deviation from thermal equilibrium, i.e.,
\begin{equation}
\bar{N}_{\vec{k}}(t) = \frac{1}{e^{\beta \bar{\omega}_k}+1} +
\delta\bar{N}_{\vec{k}}(t)~,\quad
\bar{N}_{\vec{q}}(t)=\frac{1}{e^{\beta \bar{\omega}_q}+1}~,\quad
N_{\vec{k}+\vec{q}}(t)=
\frac{1}{e^{\beta\omega_{k+q}}-1}~.
\end{equation}
In the linear relaxation approximation (or relaxation time
approximation), we find 
\begin{eqnarray}
&&\frac{d}{dt}\delta\bar{N}_{\vec{k}}(t) =-\Upsilon_k\;
\delta\bar{N}_{\vec{k}} (t)~,\label{boltzmann}
\end{eqnarray}
with $ \Upsilon_k $ being the {\em interaction rate} in the
relaxation time approximation (linear relaxation) and given by 
\begin{equation}
\Upsilon_k =
\frac{y^2 m^2 T}{16 \pi k
\bar{\omega}_k}\left(1-\frac{4M^2}{m^2}\right)
\ln\left.\left[\frac{1-e^{-\beta(\bar{\omega}_{q}+\bar{\omega}_{k})}}{1+
e^{-\beta\bar{\omega}_{q}}}\right]\right |
^{\bar{\omega}_{q^{\ast}_{2}}}_{\bar{\omega}_{q^{\ast}_{1}}}~,
\label{relaxtime} 
\end{equation}
where $q^{\ast}_{1}$ and $q^{\ast}_{2}$ are given by
eqs.~(\ref{q2star}). 
Comparing the interaction rate with the damping rate found in the previous
section (see eq.~(\ref{dampingratefin})), we provide a real time
confirmation of the result 
\begin{equation}
\Upsilon_k = 2\; \Gamma_k~. \label{relation}
\end{equation}

The kinetic analysis confirms that the damping of the fermionic
quasiparticle excitations in   
the medium is a consequence of the induced
decay of the heavy scalar. Furthermore this analysis {\em in real
time} clearly establishes the relation between the interaction rate
in the relaxation  
time approximation and the exponential decay of the amplitude of
the mean field at least to lowest order. Recently a detailed
investigation between the damping  
and interaction rates for chiral fermions in gauge theories at finite
temperature  
has been reported in reference~\cite{dolivodamp2} within the
framework of imaginary time (Matsubara) finite temperature field theory. 
Our analysis provides a complementary confirmation of this result in 
{\em real time} both for the relaxation of the mean field and that of the 
spin-averaged distribution function. 

\section{Conclusions, comments and further questions}

In this article we have focused on studying the propagation of fermionic
excitations in a fermion-scalar plasma, as a complement to the more studied
issue of propagation in a gauge plasma. 

Our motivation was to provide a real time analysis of the propagation that could 
eventually be used in other problems such as, for example, neutrino
oscillations in medium and in non-equilibrium processes in electroweak baryogenesis. 
The first step of the program is to obtain the effective
Dirac equation in medium and in real time. This is achieved by relating the
problem of linear response to an initial value problem for the mean
field that is  
induced by an external Grassmann-valued source term. The resulting
Dirac equation for the mean field is fully renormalized and causal and
allows a direct  
study of real time phenomena. 

We used this description to study the propagation of fermions in a 
fermion-scalar plasma to lowest order in the Yukawa coupling. We found
that when the  
scalar mass is large enough that its  decay  into fermion pairs is
kinematically allowed,  
this process in the medium leads to a {\em damping} of the fermionic
excitations and a  
quasiparticle picture of its propagation in the medium. A real time
description of this  
process clearly leads to the identification of the damping rate,
which we computed  
to one loop order for arbitrary values of the fermion and scalar
masses (provided the 
scalar is heavy enough to decay), temperature and fermion momentum. 

An all-order expression for the damping rate (in the narrow width
approximation)  is obtained from the {\em exact} quasiparticle solutions
to the in-medium Dirac equation [see eq.~(\ref{exactwidth})]. 
A kinetic approach based on a Boltzmann equation for the spin-averaged
fermionic distribution function reveals that
the interaction rate in the relaxation time approximation 
(linear departures from equilibrium) are simply related to the damping
rate of the mean fields at least to lowest order in the Yukawa coupling. 
We emphasize that this relation is established here from the {\em real time} evolutions both 
of the mean  field and the distribution function.

{\em Comments.} Although the expression for the rate was obtained for
arbitrary values  
of the scalar and fermion masses, temperature and fermion momentum, a
deeper analysis  
is required if the theory undergoes a second or
very weakly first order transition. The reason being that if the
fermionic masses are  
a result of spontaneous symmetry breaking in the scalar sector,
near a second order (or very weak first order) phase transition both
the scalar mass and  
the chiral breaking fermion mass vanish. In this case the kinematic region
in momenta for which the energy conserving delta functions are
fulfilled shrinks  
and one must understand if, for soft fermionic momentum, a resummation
akin to hard thermal loops is required. 
In particular from the expression of the spectral densities given by
eq.~(\ref{rhos}), 
it is straightforward 
to see that both $\rho^{(1)}_{\vec{k}}(k_0)$ and $\rho^{(2)}_{\vec{k}}(k_0)$
contribute in the hard thermal loop limit for vanishing scalar and
fermion mass   
very similarly to the Landau damping contribution from gauge fields~\cite{htl} 
with a cut discontinuity for space like momenta ($\omega^2< k^2$). 
This particular case, corresponding to the limit
$T\gg m,M$, would have to be studied separately and is beyond the
scope of this article.  
A possibility that arises in this situation is that as the phase
transition is approached the effective temperature dependent scalar mass
becomes smaller than twice the temperature dependent fermionic mass,
the scalar decay  
channel shuts-off and the contribution to the fermionic damping rate
vanishes. This 
of course depends on the self-couplings of the scalar sector and
requires studying in  
detail particular models. 

{\em Further questions.} 
In a full fermion, gauge plus scalar theory with light fermions of mass 
$M\ll eT$ ($e$ is the gauge coupling), 
the contribution of the gauge sector to the fermion self-energy requires the
hard thermal loop resummation for soft fermion momenta $k$ of order $eT$, 
but is perturbative for hard fermion momenta $k$ of order $T$. 
For the lightest quarks $u$, $d$ and $s$ but certainly {\em not} $c$, $b$ and $t$ 
(here we have in mind the electroweak theory and the problem of
baryogenesis), the Yukawa couplings $y\ll e$ 
and the contribution from the scalars is perturbatively small.
This aspect notwithstanding, let us consider the case of soft fermionic
momenta. The hard thermal loop resummation leads to the dispersion relations
for {\em plasmino} quasiparticles in the medium, which in lowest order
in HTL are stable collective excitations. However, with the coupling
of scalars, 
the heavy scalar (with mass of order $T$) can now decay into a soft
plasmino and a  
hard fermion (as the hard plasminos are almost indistinguishable from
the zero-temperature  
fermions). The results of this article indicate that this process will
then lead to  
a damping rate for the plasminos in lowest order (and certainly perturbative) 
in the Yukawa coupling. This process will compete with the damping of soft 
fermions through the exchange of magnetic photons, and it requires a 
detailed analysis which is currently under study~\cite{plasmino}.

\section{Acknowledgments}
D.B. and S.-Y.W. thank the NSF for partial support through grant
PHY-9605186.  D. B. and H.J.d.V. acknowledge support from NATO. 
D.-S.L. would like to thank the members of the Department of Physics and
Astronomy of the University of Pittsburgh and the Department of Physics and
Astronomy of the University of North Carolina at Chapel Hill for their
hospitality where part of this work was done. He also thanks the support
from the Republic of China National Science Council through grant NSC
88-2112-M-259-001. 
Y.J.N. thanks DOE for partial support through grant DE-FG05-85ER-40219
Task A.

%\input{fermionbiblio}

%%%%%%%%%%%%%%begin bibliography %%%%%%%%%%%%%%

%%%%%%%%%end bibliography %%%%%%%%%%%%%%%

\newpage

%%%%%%%%%%%%begin figures %%%%%%%%%%%%%%%%%%%%
%%%%figure 1 %%%%%%%%%%%%%%%%%%
\begin{center}
\begin{figure}[t] 
\epsfig{file=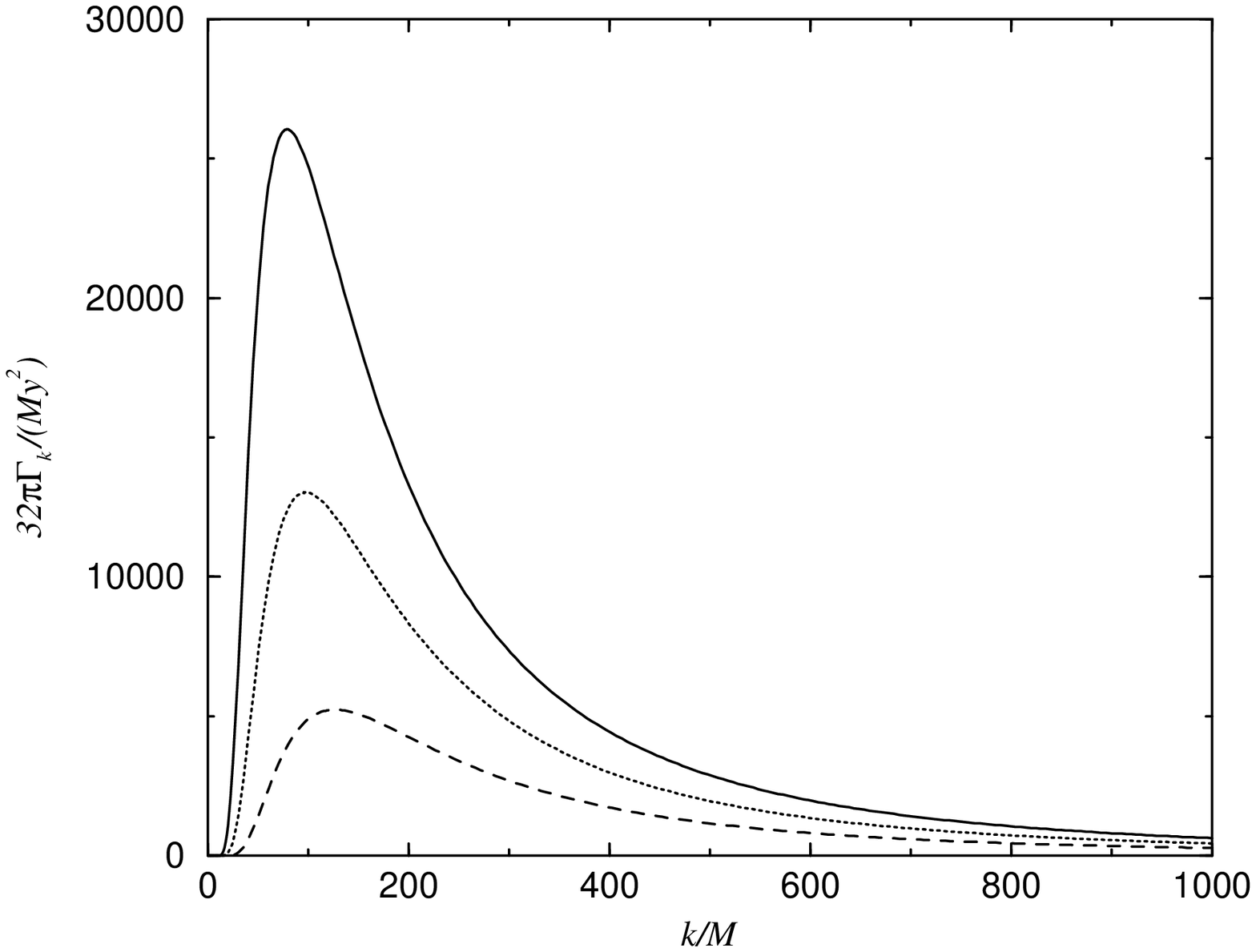,width=15cm,height=15cm} 
\caption{$32\pi \Gamma_k/(My^2)$ v.s. $k/M$ for
$m/M=800$ and $T/M= 1000$ (solid line), $800$ (dotted line), 
$600$ (dashed line). \label{fig1}}
\end{figure} 
%%%%%%%%%end figure 1
%%%%%%%%figure 2%%%%%%%%%
\begin{figure}[t] 
\epsfig{file=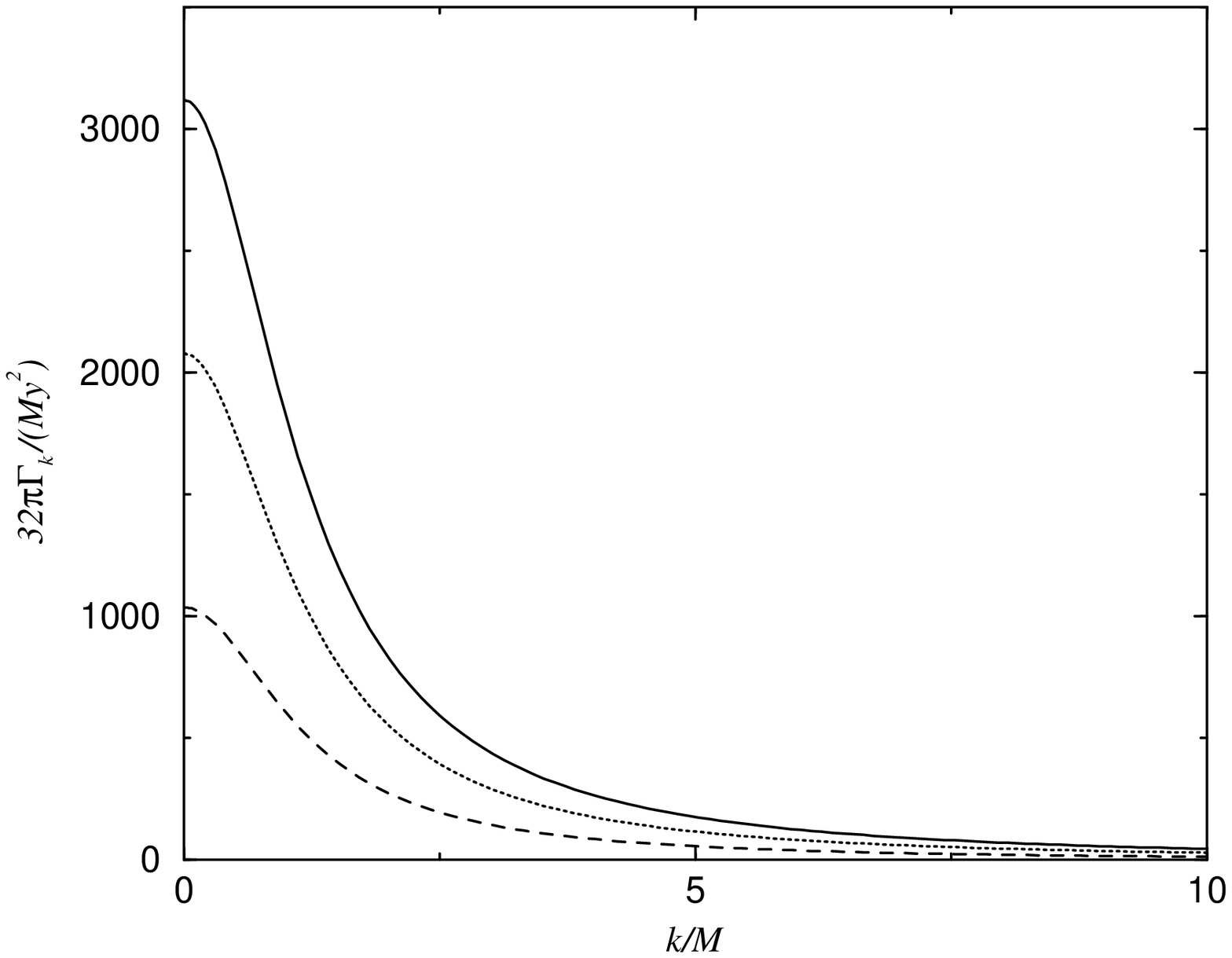,width=15cm,height=15cm} 
\caption{$32\pi \Gamma_k/(My^2)$ v.s. $k/M$ for
$m/M=4$ and $T/M=150$ (solid line), $100$ (dotted line), $50$ (dashed line). \label{fig2}}
\end{figure} 
%%%%%%%%end figure 2%%%%%%%%%
%%%%%%%%figure 3%%%%%%%%%
\begin{figure}[t] 
\epsfig{file=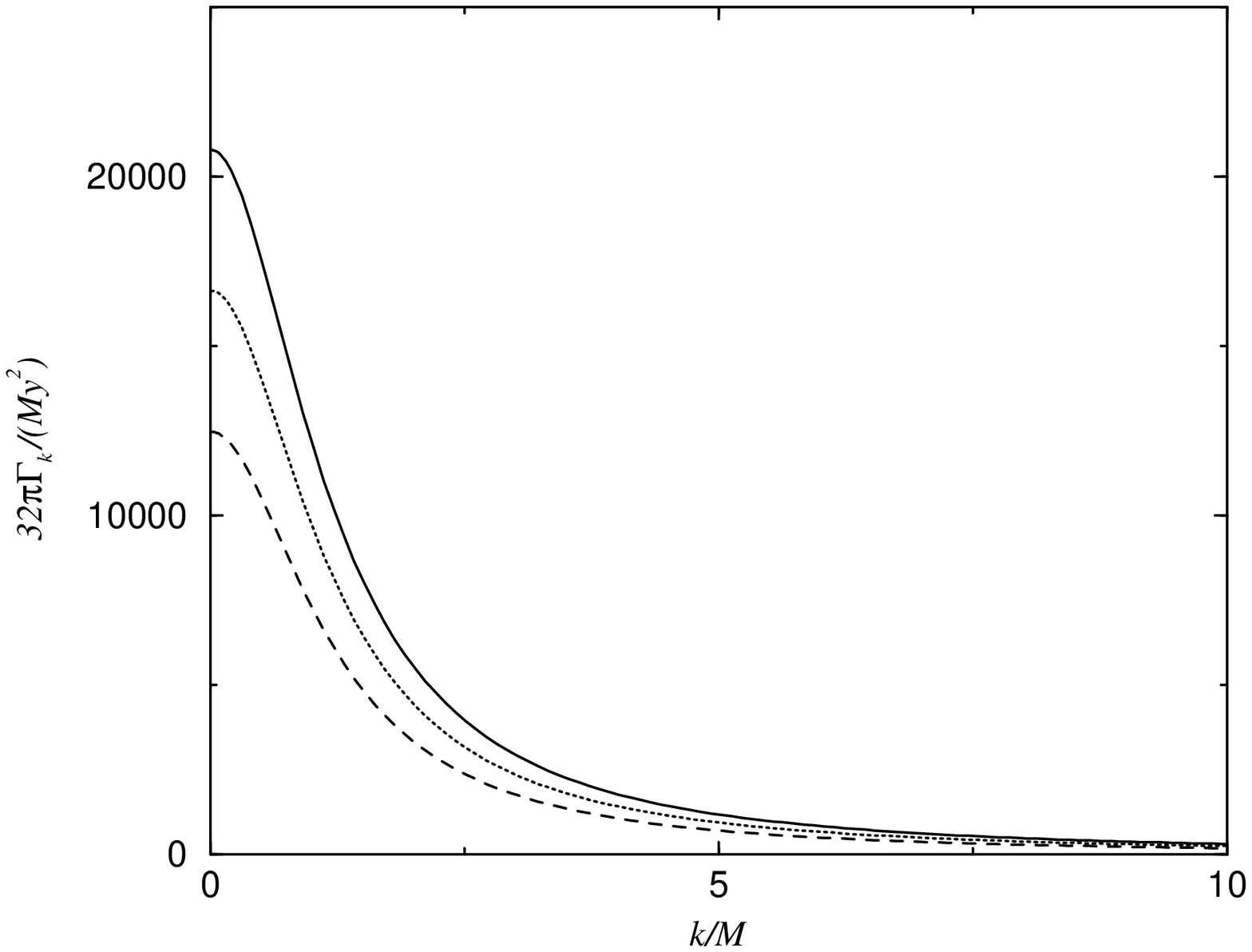,width=15cm,height=15cm} 
\caption{$32\pi \Gamma_k/(My^2)$ v.s. $k/M$ for
$m/M=4$ and  $T/M= 1000$ (solid line), $800$ (dotted line), $600$ (dashed line). \label{fig3}}
\end{figure} 
\end{center}
%%%%%%%%end figure 3%%%%%%%%%

%%%%%%%%%%%%%end figures %%%%%%%%%%%%%%%%%%%%%

\end{document}